\documentclass[runningheads]{llncs}
\pdfoutput=1
\usepackage{graphicx}
\usepackage{booktabs}
\usepackage{array}
\usepackage{amsmath,amssymb}
\usepackage{algorithm}
\usepackage[noend]{algpseudocode}
\DeclareMathOperator{\E}{\mathbb{E}}

\makeatletter
\def\BState{\State\hskip-\ALG@thistlm}
\makeatother

\begin{document}
\title{Adversarial Pulmonary Pathology Translation for Pairwise Chest X-ray Data Augmentation\thanks{This research was supported by the Australian Research Council Centre of Excellence for Robotic Vision (project number CE140100016).}}

\titlerunning{Adversarial Pulmonary Pathology Translation for Data Augmentation}

\newcommand*\samethanks[1][\value{footnote}]{\footnotemark[#1]}
\author{Yunyan Xing\inst{1} \thanks{Corresponding and equal contribution authors.} \and
Zongyuan Ge\inst{1} \samethanks\and
Rui Zeng\inst{1}\and
Dwarikanath Mahapatra\inst{2} \and
Jarrel Seah\inst{1}\and
Meng Law\inst{1}\and
Tom Drummond\inst{1}}

\authorrunning{Y.Xing, Z.Ge, et al.}

\institute{Monash University, Australia \and
IBM Research, Australia \\
\email{\{yunyan.xing, zongyuan.ge\}@monash.edu}}
\maketitle 

\begin{abstract}
Recent works show that Generative Adversarial Networks (GANs) can be successfully applied to chest X-ray data augmentation for lung disease recognition. However, the implausible and distorted pathology features generated from the less than perfect generator may lead to wrong clinical decisions. Why not keep the original pathology region? 
We proposed a novel approach that allows our generative model to generate high quality plausible images that contain undistorted pathology areas. The main idea is to design a training scheme based on an image-to-image translation network to introduce variations of new lung features around the pathology ground-truth area. Moreover, our model is able to leverage both annotated disease images and unannotated healthy lung images for the purpose of generation.  
We demonstrate the effectiveness of our model on two tasks: (i) we invite certified radiologists to assess the quality of the generated synthetic images against real and other state-of-the-art generative models, and (ii) data augmentation to improve the performance of disease localisation.

\end{abstract}
\section{Introduction}
Chest X-ray images are the most commonly used method for lung disease diagnosis. Nowadays, deep learning neural network models are available for investigation of classification, segmentation and other problems in medical imaging~\cite{esteva2017dermatologist,eaton2018improving,salehinejad2018generalization}. With the rapid growth in the number of X-ray images needed to be reviewed by radiologists in the hospital, building deep learning models as computer-aided diagnosis tools that quickly and accurately detect the pathology area can help reduce the workload of radiologists. 
Training a robust deep learning model requires a large number of high quality samples with accurate pathology information. 
However, it is difficult to obtain enough medical images because many diseases are rare.

Generative Adversarial Networks (GANs)~\cite{goodfellow2014generative} have become a new technique to perform data augmentation. Compared with the traditional mathematical data augmentation methods, such as rotations, translations, reflections and adding Gaussian noise~\cite{krizhevsky2012imagenet}, the advantage of GANs is that the models can generate new synthetic data with much larger diversity. Various GAN models have been proposed to generate synthetic images for data augmentation.  
Image-to-image translation models such as StarGAN~\cite{choi2018stargan} and CycleGAN~\cite{zhu2017unpaired} have been applied on CelebA~\cite{liu2015deep} to generate new face images. 
DCGAN~\cite{radford2015unsupervised} has been applied in the chest X-ray imaging domain to perform data augmentation, showing improvement in disease classification accuracy~\cite{salehinejad2018generalization,frid2018synthetic}.

Although the discussed data augmentation methods can improve recognition performance for some tasks, it is difficult to explain whether the augmented images benefit the model regularisation or discrimination. This is because medical images generated by DCGAN do not look visually plausible and vital disease features are missing. This would not be an issue for a general dataset such as ImageNet~\cite{deng2009imagenet} or a face dataset such as CelebA, because deep learning models can make correct decisions based on the overall and contextual features. However, in the medical domain, clear and distinguishable biomarkers and pathological evidence are crucial to driving clinical decisions. 
To the best of our knowledge, none of the aforementioned GAN models can fully retain undistorted pathology areas for newly generated images.

To address the above issues, we employ an idea similar to image inpainting by generating regions around the undistorted pathology area with a pairwise image-to-image translation (Pix2Pix) model~\cite{isola2017image}. 
We leverage the limited number of bounding box labelled pathology data and unlabelled healthy data from the NIH Chest X-ray dataset~\cite{wang2017chestx} to create a set of artificial paired training images to fit into the training process of the Pix2Pix model. Our proposed approach is shown in Fig.~\ref{fig1}.
The generated chest X-ray images from our proposed method contain visually plausible lung structures and pathology evidence. We validate and show the superior performance of our model through two experiments: (i) a radiologist assessment test and (ii) data augmentation for lung disease localisation.

\section{Methodology}
\subsection{Image-to-Image Translation with GANs}
Recent work in generative adversarial networks (GANs) have achieved impressive results on image generation by using image-to-image translation~\cite{choi2018stargan,isola2017image}.
Image-to-image translation learns a mapping from an input image $x$ to a target output image $y$.
Our target is to generate new disease training samples $\tilde{x}$ stemming from the original samples $x$ via the image-to-image translation framework. 
In this work, we use Pix2Pix GAN as our main image generation framework.  

\begin{figure}
\includegraphics[width=\textwidth]{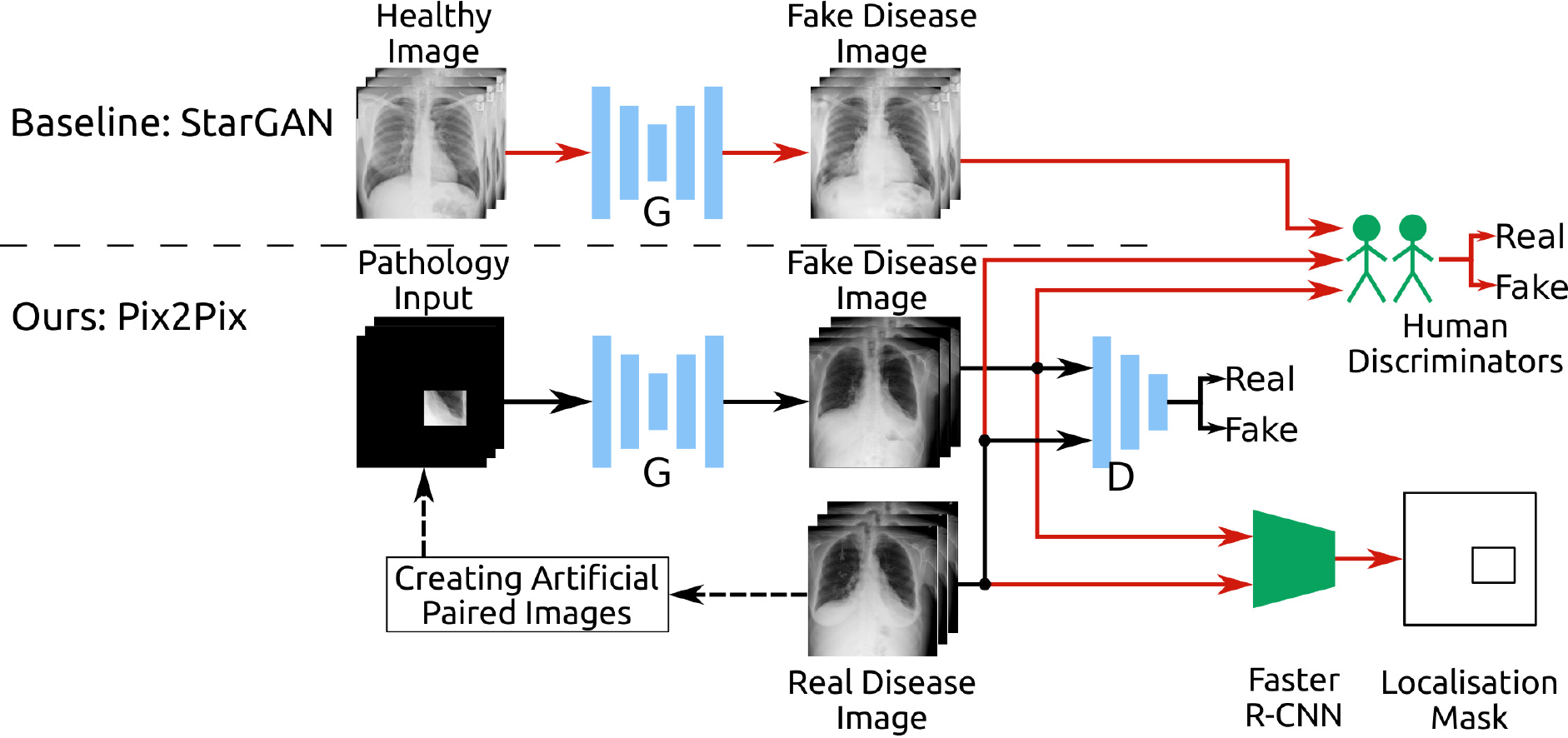}
\caption{The black arrows indicate the training process: for StarGAN (top), which is our baseline, the generator is given unpaired chest X-ray images to perform healthy image to diseased image translation.
For our proposed Pix2Pix model (bottom), the generator is given an input image with the masked pathology area and is trained to produce a complete X-ray image with the surrounding lung details.
The red arrows indicate the evaluation process: we invite two radiologists to act as `human discriminators' to assess the quality of our synthetic images. We also use a Faster R-CNN model to measure the performance of our approach on data augmentation for disease localisation.
} \label{fig1}
\end{figure}

\noindent\textbf{Pix2Pix GAN Loss.} The loss function for Pix2Pix consists of two parts. The first part is the adversarial loss, which is for the discriminator $D$ to ensure the synthetic image $\tilde{x}=G(x)$ generated by the generator can be distinguished from the original target real image $y$. Further, the generator $G$ is trained to minimise the adversarial loss to make the synthetic image more plausible. Therefore, $D$ and $G$ update iteratively against each other with the following objective function:
\begin{equation}
L_{GAN}(G, D) = \E_{x,y}[\log D(x,y)]+\E_{x}[\log (1-D(x,G(x)))],
\end{equation}
where G is trying to minimise this objective against D, which tries to maximise it.~\cite{isola2017image}. In addition, the Pix2Pix model includes a traditional construction loss, L1 distance, to ensure the quality of generated synthetic images:
\begin{equation}
L_{L1}(G)=\E_{x,y}[||y-G(x)||_1].
\end{equation}
Therefore, the full objective function for Pix2Pix is:
\begin{equation}
G^\ast, D^\ast= \arg \min_G \max_D L_{GAN}(G,D)+\lambda L_{L1}(G).
\end{equation}

\subsection{Pulmonary Pathology Data Augmentation}
The key feature that makes the Pix2Pix model powerful in generating high quality synthetic images is that Pix2Pix requires paired images to learn the mapping from input to output images~\cite{isola2017image}.
However, it is almost impossible to obtain many chest X-ray training pairs with the same patient ID and perfect alignment of the two lungs, but containing two different conditions. To address this issue, we design an image in-painting method to craft artificial paired training images using the limited number (820 across 6 diseases) of bounding box annotated images from the NIH dataset~\cite{wang2017chestx}. 
Our idea is simple but effective: we let Pix2Pix learn how to recover the missing regions around the known pathology area with visually plausible image structures and textures. Each iteration, the model will generate new synthetic images with the same undistorted pathology area, which is vital to diagnose or localise pulmonary diseases, but with variations in the surrounding area. Therefore, the Pix2Pix generated images can be used as extra training samples to augment the training data. 

\noindent\textbf{Creating Artificial Pairwise Training Images.}
We take each of the annotated images $y \in [1,...,N]$ as the the target output images and create a `pathology area only' image $x$ with binary image mask $m$. Input image $x$ is constructed from the raw images as $x \leftarrow y \odot m$. Mask $m$ replaces the area outside the bounding box with zero values, keeping the area inside the mask unchanged. We use this `pathology area only' image $x$ and the corresponding original image $y$ to create a training pair $(x,y)$. This allows us to perform image-to-image translation on the pairwise images. 

\noindent\textbf{Training Pix2Pix with Artificial Training Pairs.}
During training, we pre-process the data to prepare training pairs $(x,y)$. Generator $G$ takes $x$ as input, and outputs predicted image $x^{'}=G(x)$ with the same size as the input. Combining image $x^{'}$ with the input image $x$, we get the new output $\tilde{x} \leftarrow x + G(x) \odot (1-m)$. Then we train the discriminator with the adversarial loss to try to distinguish between the generated image $\tilde{x}$ and the real image $y$. We step the gradient descent algorithm on the discriminator and generator alternatively. The training procedure is shown in Algorithm~\ref{euclid}.  
This training scheme means that the vital pulmonary pathology features from the original image bounding box are undistorted and well-retained in the generated synthetic images. This lowers the difficulty of the generator training. After employing our training technique, the generator only needs to learn how to reconstruct the surrounding lung features outside of the pathology area.

\begin{algorithm}
\caption{Training of pulmonary pathology data augmentation.}\label{euclid}
\begin{algorithmic}[1]
\While{G has not converged}
    \For{$i=1,...N$}
        \State Sample target images $y_{i}$ from the data with bounding box annotation; \label{batch_sample}
        \State Extract masks $m_{i}$ for $y_{i}$ according to annotations;
        \State Construct inputs $x_{i} \leftarrow y_{i} \odot m_{i}$;
        \State Craft training pairs $(x_{i},y_{i})$;
        \State Get predictions $\tilde{x_{i}} \leftarrow x_{i} + G(x_{i}) \odot (1-m_{i})$;
        \State Calculate gradients for $G$ and $D$;
        \State Update $G,D$;
    \EndFor 
\EndWhile
\end{algorithmic}
\end{algorithm}

\noindent\textbf{Adding Unannotated Healthy Samples.}
The primary goal of our Pix2Pix model is to retain undistorted pathology area and to train the generative model only to learn how to generate the lung features. To further extend the number of training samples, we discover that adding normal healthy lung X-ray images with random cropping areas into the training set can increase the quality of the synthetic images. 
By following the same training procedure in Algorithm 1, we add an extra 2,000 healthy lung X-ray images $\hat{y}$ and randomly generate masks $\hat{m}$ with reasonable size. Healthy pairwise images $(\hat{x} \leftarrow \hat{y} \odot \hat{m}, \hat{y})$ are sampled along with disease images in Line~\ref{batch_sample} of Algorithm 1. 

\noindent\textbf{Comparing with StarGAN.}
In this paper, we compare our proposed generative method to another image-to-image translation GAN, StarGAN~\cite{choi2018stargan}, which is based on CycleGAN~\cite{zhu2017unpaired}. The main advantage of either CycleGAN or StarGAN is that they can be trained using unpaired images. Bounding box annotations are not required to translate an image with a given disease label into other disease labels. The unique features belonging to a certain category are supposed to be learned implicitly during model training. 
Since the generator $G$ and discriminator $D$ are less than perfect, the structure and texture information of some pulmonary disease translations are not always correct, especially for diseases that have high intra-class variations, such as Pneumonia.   
This approach works well on tasks such as face generation; people are capable of recognising the identity in the generated image even with some misleading contents. However, this may lead to a catastrophic outcome when a deadly disease can not be recognised because of blurry or inaccurate generated pathology evidences. 
This is the reason we propose the generating mechanism to keep the most vital pathology information undistorted. The pairwise training requirement of PixPix guarantees the `upper bound' that CycleGAN can achieve~\cite{zhu2017unpaired}. 
The training and evaluation procedures for our Pix2Pix model are shown in Fig.~\ref{fig1}.

\begin{figure}[!t]
\includegraphics[width=\textwidth]{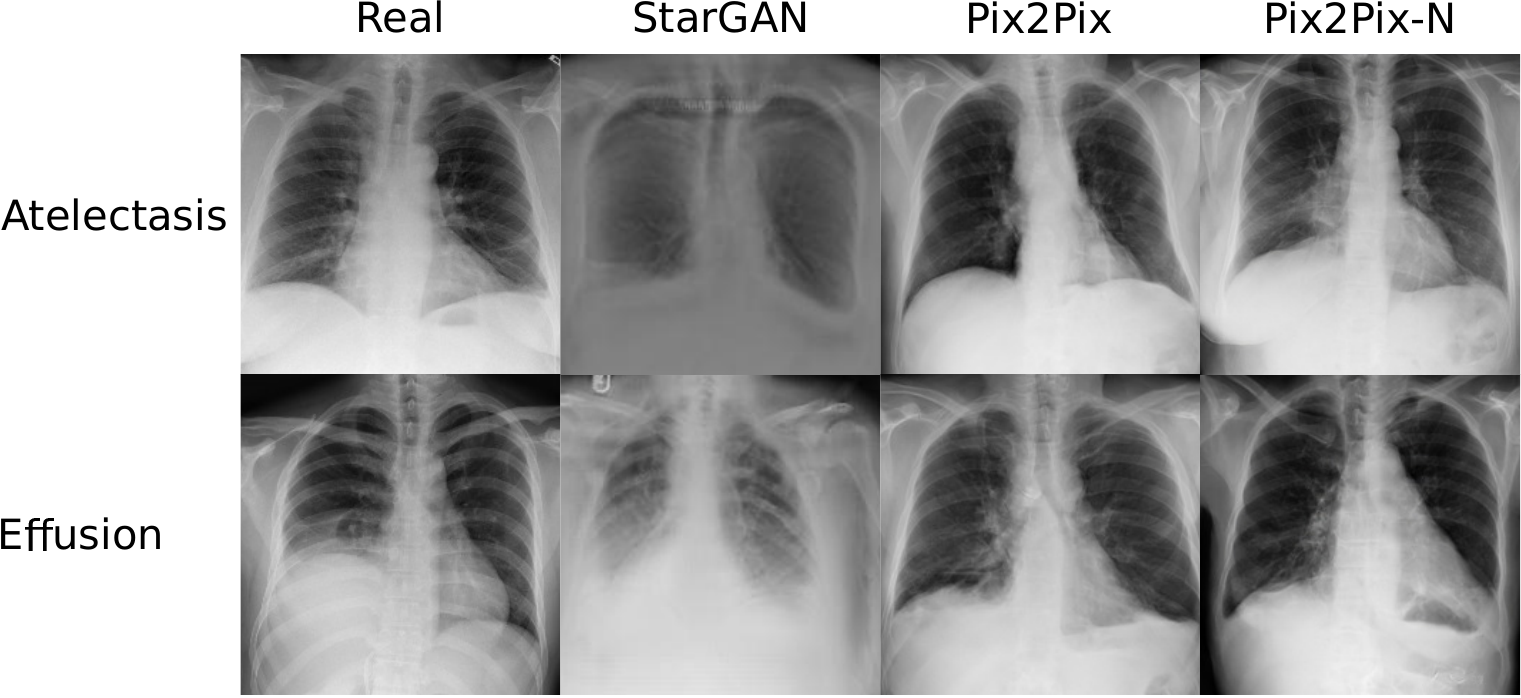}
\caption{Samples of real images and synthetic images generated by StarGAN and our models (Pix2Pix and Pix2Pix-N).} \label{fig2}
\end{figure}

\section{Experiments}
\noindent\textbf{Dataset.} We use the NIH Chest X-ray dataset~\cite{wang2017chestx}, which is by far the largest publicly available chest X-ray dataset. It consists of 112,120 frontal-view chest X-ray images with 14 disease categories and the healthy category. Each image is labelled with one or multiple diseases. The dataset also contains 984 bounding box annotations for 880 images corresponding to eight disease categories. We use the bounding box annotated images to create a training set for our Pix2Pix model. However, for two diseases among those eight disease categories, the number of bounding box annotations is too little and is not sufficient to train the Pix2Pix model. As such, we remove those two diseases from our training set. The final training set we use for our model contains 820 bounding box annotations for 741 images over six pathologies.

\noindent\textbf{Implementation.} We train two Pix2Pix models that use different training sets. We use 70\% of the original 820 bounding box annotations as the training set for our Pix2Pix models and localisation model, amounting to 573 training annotations. The remaining 30\% of annotations are used as the testing set for the localisation model.
Our first model, \textbf{Pix2Pix}, is trained with only the 573 annotated images. Our second model, \textbf{Pix2Pix-N}, is trained with the 573 annotated images and 2,000 healthy images with randomly generated bounding boxes. The sample images generated by each model are shown in Fig.~\ref{fig2}. We downsize the original images from 1024 $\times$ 1024 to 256 $\times$ 256 for fast processing. We also apply mathematical transformations, including rotations, reflections and cropping, to add variation into the network.

\subsection{Qualitative Analysis}
We invite two clinicians, R1 and R2\footnote{R1 is a certified radiologist with 10 years of experience. R2 is a radiology registrar who also has two years of experience with deep learning. To ensure fairness, all of the real and generated images are cropped to remove most of the artefacts and downsized to $224 \times 224$. The radiologists are not aware of the disease prevalence or the proportion of real and fake images in the test set. The radiologists are asked to work independently to distinguish whether each image is real or fake.} to verify the quality of the synthetic images generated by our proposed models. 
To compare the performance of our Pix2Pix models to other generative models, we use StarGAN as a baseline. 
We provide the radiologists with a verification set contains six pathologies with 100 images each. The verification set is comprised of 150 real images from the NIH dataset and three sets of 150 images generated by each of the generative models (\textbf{StarGAN}, \textbf{Pix2Pix} and \textbf{Pix2Pix-N}).

Fig.\ref{fig2} shows qualitative illustrations and Table~\ref{radiologistsresults} shows the number of images that the radiologists identify as real from each source. 
From Table~\ref{radiologistsresults}, we observe that both R1 and R2 are good at picking real chest X-ray images. It is found that either Pix2Pix or Pix2Pix-N surpasses StarGAN by a large margin ($15|0$ vs. $108|45$). This shows: 1) the merits of keeping undistorted pathology information, and 2) Pix2Pix based models produce plausible reconstructions of the surrounding lung features.  
Pix2Pix-N performs better than Pix2Pix, showing the necessity to include more healthy images for model generalisation. 
Finally, R2 is better than R1 at recognising the generated images as fake. This is likely because R2 has two years of research experience with GANs and is good at recognising GAN artefacts.

{
\setlength{\tabcolsep}{4pt}
\begin{table}[!t]
\centering
\caption{Radiologist assessment results. The two values for each element are the number of images predicted as real by R1$|$R2, respectively. For the generative models (StarGAN, Pix2Pix and Pix2Pix-N), larger values indicate better performance. }\label{radiologistsresults}
\begin{tabular}{lcccc}
\toprule
Pathology & Real Data & StarGAN & Pix2Pix & Pix2Pix-N\\
\midrule
Atelectasis &  $25|25$ & $2|0$ & $12|2~ $ & $19|6~ $ \\
Cardiomegaly &  $25|25$ & $2|0$ & $18|4~ $ & $21|8~ $ \\
Effustion &  $25|24$ & $5|0$ & $11|3~ $& $17|6~ $ \\
Infiltration &  $25|25$ & $6|0$ & $11|3~ $ & $18|10$\\
Pneumonia &  $25|25$ & $0|0$ & $14|4~ $ & $19|11$\\
Pneumothorax &  $25|25$ & $0|0$ & $13|2~ $ & $14|4~ $ \\
\midrule
Total &  150$|$149 & 15$|$0~~  & 79$|$18 & 108$|$45~ \\
\bottomrule
\end{tabular}
\end{table}
}

\subsection{Disease Localisation}
In this section, we investigate the effectiveness of synthetic images generated by our proposed model on the task of pathology localisation. 
A Faster-RCNN~\cite{he2017mask} built on InceptionV2-ResNet is used as the detection model. We train the detection model using three different dataset augmentation protocols: original (\textbf{Ori}), \textbf{Ori+Pix2Pix}, and \textbf{Ori+Pix2Pix-N}. Ori uses the original images (573 in total) from the NIH dataset. Ori+Pix2Pix contains all real images and 688 synthetic images obtained from our Pix2Pix model. Ori+Pix2Pix-N is composed of all real images and 688 synthetic images obtained from our Pix2Pix-N model.
For fair comparison, all protocols use the same parameters to perform the Faster-RCNN training. Specifically, all datasets are resized to $256 \times 256$ without any further augmentation. The learning rate is 0.0003 and the batch size for each training step is 2. The evaluation dataset consists of 247 real images. To clearly observe the trends of the performance of each dataset, correct location (CL) accuracy computed at an intersection over union (IoU) of 0.1 is chosen as the evaluation metric, as per \cite{wang2017chestx}. To reduce the chance of performance oscillation, the results reported at a given training step $s$ are selected from the best model in a step range $[s-500, s+500]$. 

{
\setlength{\tabcolsep}{2pt}
\begin{table}[!t]
	\centering
	\scriptsize
	\caption{Disease Localisation results of Ori, Ori+Pix2Pix, Ori+Pix2Pix-N datasets. $\mathrm{CL}@_{s}^{0.1}$ is the best correct location accuracy computed at an IoU of 0.1 and selected from $[s-500, s+500]$ steps.}
	\begin{tabular}{lccccccccc}\toprule
		&  \multicolumn{3}{c}{Ori}    & \multicolumn{3}{c}{Ori+Pix2Pix}   & \multicolumn{3}{c}{Ori+Pix2Pix-N}	\\
		\cmidrule(lr){2-4} \cmidrule(lr){5-7} \cmidrule(lr){8-10}
		Pathology & $\mathrm{CL}@^{0.1}_{5k}$ & $\mathrm{CL}@^{0.1}_{10k}$ & $\mathrm{CL}@^{0.1}_{15k}$ & $\mathrm{CL}@^{0.1}_{5k}$ & $\mathrm{CL}@^{0.1}_{10k}$ & $\mathrm{CL}@^{0.1}_{15k}$ & $\mathrm{CL}@^{0.1}_{5k}$ & $\mathrm{CL}@^{0.1}_{10k}$ & $\mathrm{CL}@^{0.1}_{15k}$\\
		\midrule
		Atelectasis  & 0     & 0     & 0.018 & 0     & 0     & 0     & 0     & \textbf{0.018} & \textbf{0.111} \\
		Cardiomegaly & 0.636 & 0.818 & 0.818 & 0.613 & 0.750 & 0.681 & \textbf{0.636} & \textbf{0.840} & \textbf{0.863}\\
		Effusion     & 0.173 & 0.217 & 0.304 & 0.173 & 0.195 & 0.130 & \textbf{0.239} & \textbf{0.282} & \textbf{0.326}\\
		Infiltration & 0.378 & 0.432 & 0.459 & 0.297 & 0.351 & 0.324 & \textbf{0.405} & \textbf{0.513} & \textbf{0.594}\\
		Pneumonia    & \textbf{0.333} & 0.305 & 0.305 & 0.194 & 0.250 & 0.277 & 0.166 & \textbf{0.361} & \textbf{0.527}\\
		Pneumothorax & 0.031 & 0.033 & 0.066 & 0     & 0     & 0.033 & \textbf{0.033} & \textbf{0.033} & \textbf{0.100}\\
		\midrule
		Total        & \textbf{0.259} & 0.301 & 0.328 & 0.213 & 0.234 & 0.233 & 0.235 & \textbf{0.323} & \textbf{0.405}	\\
		\bottomrule
	\end{tabular}
	\label{tab:fastercnnablation}
\end{table}
}

Table~\ref{tab:fastercnnablation} shows the disease localisation performance. The synthetic augmentation protocol \textbf{Ori+Pix2Pix-N} performs the best among all three datasets in terms of model performance and convergence. It significantly improves the disease localisation accuracy and convergence speed of the Faster-RCNN model. We conjecture that images generated from pathology annotated images and healthy images explicitly carry useful pathology information from the real data distribution, which is crucial for deep model localisation training. The \textbf{Ori+Pix2Pix} performed the worst among three training strategies. This is probably because synthetic disease images overfit the model training and increase the data bias.

\section{Conclusion}
In this paper, we proposed a model to perform data augmentation on pairwise chest X-ray images by using an image-to-image translation model (Pix2Pix). Our approach is able to generate high quality and plausible synthetic images with undistorted pathology areas, which is crucial for disease diagnosis and pathology area localisation. Our experimental results show that the synthetic images generated by our model are of a greater quality than those generated by StarGAN and can significantly improve the performance of a deep learning disease localisation model on unseen data.

\end{document}